\documentclass[prl,preprint,showpacs,preprintnumbers,amsmath,amssymb,superscriptaddress]{revtex4}

\usepackage{graphicx}
\usepackage{dcolumn}
\usepackage{bm}
\usepackage{float}
\usepackage{placeins}
\usepackage{lscape}
\usepackage{color}
\usepackage{ulem}
\usepackage{multirow}

\begin{document}
\draft
\title {Prediction of large linear-in-k spin splitting for holes in the 2D GaAs/AlAs system}
\author{Jun-Wei Luo}
\affiliation{National Renewable Energy Laboratory, Golden, Colorado
80401, USA}
\author{Athanasios N. Chantis}
\affiliation{Los Alamos National Laboratory, Los Alamos, New Mexico
87545, USA }
\author{Mark van Schilfgaarde}
\affiliation{Arizona State University, Tempe, Arizona 85287, USA}
\author{Gabriel Bester}
\affiliation{Max Planck Institute for Solid State Research, D-70569
Stuttgart, Germany}
\author{Alex Zunger}
\affiliation{National Renewable Energy Laboratory, Golden, Colorado
80401, USA}

\date{\today}

\begin{abstract}
The spin-orbit interaction generally leads to spin splitting (SS) of
electron and hole energy states in solids, a splitting that is
characterized by a scaling with the wavevector $\bf k$. Whereas for
{\it 3D bulk zincblende} solids the electron (heavy hole) SS
exhibits a cubic (linear) scaling with $k$, in {\it 2D
quantum-wells} the electron (heavy hole) SS is currently believed to
have a mostly linear (cubic) scaling. Such expectations are based on
using a small 3D envelope function basis set to describe 2D physics.
By treating instead the 2D system explicitly in a multi-band many-body approach we discover a large
linear scaling of hole states in 2D.
This scaling emerges from hole bands coupling that would be unsuspected
by the standard model that judges coupling by energy proximity. 
This discovery of a linear Dresselhaus k-scaling for holes in 2D implies a different understanding of hole-physics in low-dimensions.
\end{abstract}


\maketitle

Spin-orbit interaction causes the energy levels of 3D bulk-solids
\cite{Dresselhaus} and 2D quantum wells (QW's) \cite{Rashba84} to
exhibit a zero-field SS for sufficiently low-symmetry
states. On the experimental side, attention has recently focused on
spin of {\it holes} in 2D quantum-wells because of their spin
Hall-effect \cite{Murakami} and in 0D quantum dots because of the
highly coherent hole spin \cite{Brunner_Science} and unusually long
hole spin lifetime \cite{Heiss_PRB, Yogova_PRL}, promising potential
interesting applications in spintronic devices and solid state
quantum computers. On the theoretical side, the long-standing
tradition \cite{Dresselhaus, Rashba84, ZMbook} has been to describe
hole or electron spin physics in low-dimensional ($<$ 3D)
nanostructures by an expansion in a rather small basis of 3D bulk
envelope functions, using effective-mass approaches. In general, when
the basis set is restricted, the resolution of the expansion is
limited. Such low-resolution expansions can be ``far sighted''
\cite{AZ_pss} in that the actual atomistic symmetry of the nano
object \cite{Magri_PRB01} is replaced by a fictitious higher
symmetry, thus missing important degeneracy-splitting and inter-band
coupling effects. The farsightedness can be cured by systematically
increasing the basis set \cite{Cardona_PRB88} or by introducing {\it
ad-hoc} terms in the Hamiltonian intended to lower the symmetry
\cite{Foreman_PRL01, Ivchenko_PRB}. Both modifications come at the
expense of introducing more parameters that are not calculable by
the envelope function theory itself. Indeed, in the standard model
for SS of nanostructures \cite{Rashba88, Winkler_book,
Bulaev_PRL}, one uses a phenomenological Hamiltonian where one needs
to decide at the outset, which 3D bands couple in 2D by the
spin-orbit interaction (SOI), rather than have the theory force such
realization upon us. The potential of missing important physical
interactions not selected to be present in the model Hamiltonian can
be substantial \cite{AZ_pss}.

The current state of the art for the hole states in 2D is
illustrated by the work of Bulaev and Loss \cite{Bulaev_PRL}.
Starting from a bulk 3D Hamiltonian restricted to heavy-hole (HH) and
light-hole (LH) bands (``$4\times 4$"), they have derived an
effective $2\times 2$ Hamiltonian for the 2D heavy hole (hh0)
subband, demonstrating an exact cancellation of the linear-in-k
(Dresselhaus) terms \cite{Bulaev_PRL}. This result (implying a pure, uncoupled hh0 state in low-dimensions) has been used in
numerous theories of hole spin in 2D quantum-well and 0D quantum-dot
systems, including in estimation of hole spin relaxation time
\cite{Heiss_PRB}, demonstration of intrinsic hole spin Hall Effect
\cite{Bernevig_PRL}, and other hole spin related phenomena
\cite{BNL_PRL07,SLB09,OWS08,SL05}. We adopt instead a different
approach in which the 2D nanostructures is viewed as a system in its
own right, rather than express it in terms of a pre-selected basis
drawn from a 3D system. We do so by solving the 2D band structure
using explicitly the microscopic potential of the 2D system under
consideration, thus freeing us from the need to judge at the outset
which selected 3D bands (e.g., $4\times 4$ in Ref.
\onlinecite{Bulaev_PRL}) will couple in 2D. The results show that
the linear term  for {\it holes} is of the same order of
magnitude as the well known linear term for {\it electrons}
\cite{Rashba88, Eppenga_PRB}. This discovery of a linear Dresselhaus k-scaling for holes in 2D implies a different understanding of hole-physics in low-dimensions.

The central point of the approach utilized here is that the 3D and
2D systems are {\it each} described by their own microscopic
Hamiltonian which is solved in basis sets whose sole property is
that it produces a converged solution to the system at hand. Thus,
the solution reflects only the underlying microscopic Hamiltonian,
unmasked by issues of choices of bases or pre-selected Hamiltonian
terms. We use a rather general microscopic Hamiltonian in the
``{$GW$} representation''
\begin{equation}
H_{GW} = -\frac{1}{2}\nabla^2 +H_{so} +
\left[V_{ext}+V_H+\Sigma\right], \label{eq:GW}
\end{equation}
where $V_{ext}$ is the electron-ion potential, $V_H$ is the
interelectronic Hartree potential of the specific (3D or 2D) system;
$\Sigma=iG^0W$ is the self-energy with $W$ being the screened Coulomb
interaction and $G^0=1/(\omega-H^0\pm i\epsilon)$ is the Green's
function of a noninteracting Hamiltonian $H^0$. The physics
represented in Eq.~(\ref{eq:GW}) includes the atomistic symmetry of
the specific (2D or 3D) system specified by atomic position vectors
in $V_{ext}({\bf R}_n-{\bf d}_{\alpha,j})$ and incorporates
self-consistent electrostatic and exchange-and-correlation effects.
This quasi-particle self-consistent $GW$ (QS$GW$) scheme has been
established as capable of predicting accurate bulk energy bands \cite{GW},
including the Dresselhaus splitting in bulk GaAs \cite{Chantis_PRL,Krich_PRL}.

The approach described above is computationally intensive and can be
readily applied only to rather small nanostructures. Thus, for
computational expediency, when considering larger period quantum
wells [e.g, (GaAs)$_n$/(AlAs)$_n$ with $n\gg 2$] we will map both
the small-n behavior and the $n = \infty$ (bulk) QS$GW$ solutions of
Eq. (1) to a screened pseudopotential Hamiltonian that captures the
former limits yet can be readily applied to orders of magnitude
larger systems ($10^6$ atoms were demonstrated in Ref.
\onlinecite{AZ_pss}):
\begin{equation}
H_{\textrm{PP}} = -\frac{1}{2}\nabla^2 + H_{so} + \sum_{n, \alpha,
j} v_{\alpha}({\bf r}-{\bf R}_{n}-{\bf d}_{\alpha,j}).
\end{equation}
Here, the external ($V_{ext}$) plus screened ($V_H+\Sigma$) terms of
Eq. (1) are described by a superposition of atom-centered functions
$v_{\alpha}$ (where ${\bf d}_{\alpha,j}$ is the position of atom $j$
of type $\alpha$ in the $n$-th cell ${\bf R}_n$). They can be
constrained to fit approximately yet accurately the QS$GW$
calculated SS of bulk solids ($n=\infty$) \cite{Luo09}
and of low-period ($n \sim 2$) 2D quantum-wells. In addition, they
reproduce well not only the bulk band gaps throughout the Brillouin
zone, but also the electron and hole effective-mass tensors, as well
as the valence band and conduction band offsets between the well and
barrier materials \cite{AZ_pss, Magri_PRB01, Luo09}. The spin
splitting $\Delta_i$ of band $i$ obtained by the direct calculation
of Eq. (1) or Eq. (2) will be fitted to the conventional form
$\Delta_i=\alpha_i k+\gamma_i k^3$ for i = electrons (e) or holes
(h), allowing both a linear in k and a cubic in k terms to be
present.

\paragraph{Results of the many-body multi-band calculation:}
For {\bf 3D bulk GaAs}, Fig. 1 compares our multi-band approach to the {\it ab initio} 
QS$GW$ for 3D bulk GaAs, showing the linear and cubic SS of the three lowest {\it conduction bands} 
(CB1, CB2, and CB3) and three highest {\it valence bands} (HH, LH, and SO).  
The coefficients of linear and cubic
terms as obtained by fitting the calculated SS to
$\Delta_i(k)$ are given for each band in the inset to Fig. 1. We
find that the SS of all bands has a cubic term
$\gamma_i^{(3D)}$ , but only HH, LH, and CB3 with angular momentum
$J=3/2$ have linear term $\alpha_i^{(3D)}$. For these bands the
cubic and linear splittings have opposite signs. The screened
pseudopotential solution of Eq. (2) gives similar results to QS$GW$
(see inset to Fig. 1). The exception is that Eq. (2) gives
$\alpha_i^{(3D)}=0$ for all bands because this approach is coreless
and this term results from coupling to the core states \cite{Cardona_PRB88}. Fortunately,
$\alpha_i^{(3D)}$ does not matter much for 2D so we can safely use
this method for larger systems.

For {\bf 2D (GaAs)$_2$/(AlAs)$_2$ superlattice}, the SS
of {\it electrons} obtained by our atomistic multi-band approach agrees
well with {\bf k$\cdot$p} \cite{Eppenga_PRB} in that the linear term
$\alpha_e^{(2D)} \propto \alpha_e^{(3D)}/d^2$ originates from the folding-in
of 3D bulk cubic term $\alpha_e^{(3D)}$ due to the confinement to
a well of width $d$. This is not the case for {\it holes}. The
SS of 2D valence subbands (hh0, lh0, and hh1) is
presented in Fig. 2; the first two lines of Table I give the linear
and cubic coefficients $\alpha^{(2D)}_{hh0}$ and
$\gamma^{(2D)}_{hh0}$.  Both atomistic multi-band methods [Eqs. (1)
and (2)] show: (i) a linear scaling of SS in addition
to a cubic scaling for all three valence subbands including hh0,
in contrast to only a cubic scaling of hh0 in the
model-Hamiltonian derived by Bulaev and Loss \cite{Bulaev_PRL}. (ii)
The linear term dominates the SS at small k;
$\alpha_{hh0}^{(2D)}$, $\alpha_{hh1}^{(2D)}$, and
$\alpha_{lh0}^{(2D)}$ are comparable.

We next consider a few possible scenarios that might have led to a strong 2D hole splitting, finding all but the last to be unlikely.

\paragraph{(i) The standard $2\times 2$ model Hamiltonian for 2D does not explain the results.}
Bulaev and Loss \cite{Bulaev_PRL} have shown that starting from a
$4\times 4$ basis in 3D there is no linear terms  for hh0 in the 2D model Hamiltonian. However, they did not include the {\it 3D relativistic
cubic terms} of $\Gamma_{8v}$ bands in their derivation. Rashba and Sherman
\cite{Rashba88} demonstrated earlier that the 3D relativistic
cubic terms can give rise to 2D linear term for hh0 subbands [Eq. (8)
in Ref. \onlinecite{Rashba88}]. 
We tested this idea calculating the 2D SS using a pseudopotential which was constructed to fit the Rashba-Sherman 3D
band structure including the relativistic cubic terms. The results of the pseudopotential calculation for 2D (in which all bands are allowed to couple) are compared with the Rashba-Sherman 2D model (in which the 2D hh0 band is uncoupled).
Fig. 3 and Table I show that for 
sufficiently large superlattice periods $n$, for which the 2D model of Ref.~\onlinecite{Rashba88} is applicable, the model recovers only a small fraction of the multi-band results for linear SS of 2D hh0.
For smaller periods ($n \le 20$) we report in Fig. 3 a clear non-monotonic
period dependence of both $\alpha_{hh0}^{(2D)}$ and
$\alpha_{lh0}^{(2D)}$, with $\alpha_{hh0}^{(2D)}$ to
$\alpha_{lh0}^{(2D)}$ ratios varying from $\sim 10$ to $< 1$ as
the period decreases from 50 to 2 ML. This is in sharp contrast with the predictions of 
the model Hamiltonian \cite{Rashba88} which predicts a monotonic increase of linear terms
and a ratio of $\alpha_{hh0}^{(2D)}$ to $\alpha_{lh0}^{(2D)}$ which is independent of $d$.

\paragraph{(ii) Interfaces induce only minor 2D linear splitting:}
In 2D quantum-wells, Foreman \cite{Foreman_PRL01} suggested an
interface induced linear term that originates from valence band
coupling to the $p$-like $\Gamma_4$ states. For 2D GaAs/AlAs quantum
well,  he \cite{Foreman_PRL01} estimated that independent of period this interfacial
linear term is in the range of $20-30$ meV{\AA}. In contrast, 
we find for near-bulk (large-period)
superlattice such as (GaAs)$_{80}$/(AlAs)$_{20}$ (Table I) that
$\alpha_{hh0}^{(2D)}$ is very close to its bulk value of QS$GW$.
Thus, even if we assume that the interfacial linear term is solely
responsible for the remaining discrepancy from bulk value, it is
even smaller than what was estimated by Foreman
\cite{Foreman_PRL01}. We conclude that the interfacial linear
term, if it exists, is only a minor contribution to
$\alpha_{hh0}^{(2D)}$.

\paragraph{(iii) Core-valence coupling is not the reason either:}
One might have suspected that quantum confinement pushes the valence
bands closer to core levels and thus increases the linear term due
to increased coupling. This too can be excluded since our screened
pseudopotential is a coreless method, yet we find similar results as
the all-electron QS$GW$ calculation.

\paragraph{(iv) Undiscovered spin-orbit linear terms:}
The quantitative and qualitative disagreements of atomistic
multi-band calculation with the standard model Hamiltonian approach
suggest possible undiscovered SOI terms, which are not included in
the model Hamiltonian. Such terms due to symmetry lowering down from
3D $T_d$ bulk symmetry to 2D $D_{2d}$ quantum well symmetry could
originate from coupling of 3D bands via the 2D potential and SOI.
Such coupling is signaled by the distinctly nonparabolic 2D energy
dispersion curves manifesting clear anti-crossings between
neighboring subbands, multi-band calculation displayed in the inset to Fig. 3.
In the model Hamiltonian approach \cite{Rashba88, Bulaev_PRL} the hh0, hh1, ..., wavefunctions near zone-center in 2D all derive from a {\it single} bulk state $|\textrm{HH}\rangle$ and similarly, all lh0, lh1 ..., wavefunctions in 2D derive from a {\it single} bulk state $|\textrm{LH}\rangle$.  However, the inset to Fig. 3 shows that the lh0 level lies {\it between} the hh0 and hh1 levels in 2D for all multi-band calculated GaAs periods. Clearly, the coupling of 3D states and its effects on SS of 2D bands can not be ignored as done in the model Hamiltonian approach for extremely small-k range \cite{Rashba88, Bulaev_PRL}. A better approach than this ``band decoupled'' model Hamiltonian \cite{Rashba88, Bulaev_PRL} approximation allows the 2D state hh0 to derive from a few bulk states. In such a ``mixing of decoupled states'' approximation, 
\begin{equation}
 \Psi_{hh0}^{(2D)}=w_{hh0}^{(2D)}(\textrm{HH})|\textrm{HH}\rangle + w_{hh0}^{(2D)}(\textrm{LH})|\textrm{LH}\rangle+ \cdots,
\end{equation}
where $w_{hh0}^{(2D)}(\lambda)$ is the percent weight of 3D state $\lambda=\textrm{HH, LH, ...}$ in the 2D state $\Psi_{hh0}^{(2D)}$. In the band decoupled model $w_{hh0}^{(2D)}(\lambda)\equiv0$ for $\lambda \ne \textrm{HH}$. We have calculated the weights by numerical projection of the 2D pseudopotential wavefunctions and show them in Fig. 4. In striking contrast to the assumption made in model Hamiltonian approach, the interband coupling is large even at zone center. We see that for long-period (GaAs thickness $\geqslant 20$ ML) $\Psi_{hh0}^{(2D)}$ is made of 90\% $|\textrm{HH}\rangle$ and 5\% $|\textrm{LH}\rangle$, but for shorter periods the mixing {\it increases}: the HH content drops to $\sim 70-80$ \% and the LH content raises to $10-20$ \%. This monotonicly enhanced mixing of LH and HH into the 2D hh0 as the QW period is reduced signals the breakdown of the model Hamiltonian thinking that neglects such mixing on the ground that the energy splitting of hh0-lh0 must be larger than that of hh0-hh1 for sufficient small periods \cite{Bulaev_PRL}. 

The linear coefficient of 2D hh0 SS can be written in terms of the weights in Eq. (3) in a model of ``mixing of decoupled states'' as 
\begin{equation}
 \alpha_{hh0}^{(2D)}=-w_{hh0}^{(2D)}(\textrm{HH})\tilde{\alpha}_{hh0}^{(2D)} + w_{hh0}^{(2D)}(\textrm{LH})\tilde{\alpha}_{lh0}^{(2D)}+ \cdots,
\end{equation}
where $\tilde{\alpha}_{hh0}^{(2D)}(\tilde{\alpha}_{lh0}^{(2D)})$ is the contribution of a single bulk HH (or LH) band to linear SS of 2D hh0 subband, which had been derived by Rashba and Sherman \cite{Rashba88} (the negative sign accounts for band repulsion effect). The result of the first two terms in Eq. (4) is shown as open squares in Fig. 3 and are compared with the Muti-band calculated $\alpha_{hh0}^{(2D)}$ (solid circles). We see that the mixing of decoupled states [Eq. (4)] gives a much better approximation to the full calculation than the model Hamiltonian treating one decoupled band at the time (open circles in Fig.3) \cite{note1}. Thus, the mixing of bulk bands leads to a large linear SS of 2D hh0, and is unsuspected  by the standard model that judge coupling by energy proximity. 

The emergence of a large linear
term for Dresselhaus hole SS in 2D nanostructures
suggests (i) the dominance of Dresselhaus over Rashba SOI (having a
cubic term as its lowest order term) \cite{Bulaev_PRL},
(ii) a larger spin-Hall effect  \cite{Bernevig_PRL}, and (iii) an
explanation of the observed large optical anisotropy
\cite{Foreman_PRL01}. The occurrence of a larger SS of hh0 corresponding to HH-LH coupling leads to a short hole spin-relaxtion time in 2D quantum-wells  \cite{Yogova_PRL} from the D'Yakonov and Perel (DP) mechanism \cite{ZMbook}.

A.Z. thanks E. Rashba and D. Loss for helpful discussions on this
subject. Work at NREL was funded by the U.S. Department of Energy,
Office of Science, Basic Energy Science, Materials Sciences and
Engineering, under Contract No. DE-AC36-08GO28308 to NREL. 
MvS was supported by ONR, project N00014-07-1-0479.


\FloatBarrier
\newpage

\begin{table}[!hbp]
\caption[]{\label{tab:table1} The coefficients of linear and
cubic terms $\alpha^{(2D)}_i$ and $\gamma^{(2D)}_i$ for 2D
superlattice as well as 3D bulk GaAs. Here $\alpha_i^{(2D)}$ units
in meV{\AA} and $\gamma_i^{(2D)}$ in eV{\AA}$^3$. The symbol in
parentheses denotes the character of the folded-in band (bulk
$\Gamma$-like or X-like) at the Brillouin-zone center
$\overline{\Gamma}$ of the 2D system. We show results from QS$GW$
[Eq. (1)] as well as from screened pseudopotential [Eq. (2)]. The
pseudopotential used for 2D system in this Table was constructed
specifically to predict comparable bulk SS to model
Hamiltonian result \cite{Rashba88} (see text) rather than QS$GW$
values.}
\begin{tabular}{c|cc|ccc}
\hline \hline Superlattice & \multicolumn{2}{c|}{CB1}
&  \multicolumn{3}{c}{VB1} \\
(GaAs)$_n$/(AlAs)$_m$ &\hspace{1 mm}  $\alpha_e^{(2D)}$  \hspace{1
mm} &\hspace{1 mm}    $\gamma_e^{(2D)}$ \hspace{1 mm} & \hspace{1
mm} $\alpha_{hh0}^{(2D)}$ \hspace{1 mm}
& \hspace{1 mm} $\gamma_{hh0}^{(2D)}$ \\
\hline
$GW$:  2/2    ($X$)     &61.5   &0.2     &102.6     &151.3   \\
PP: 2/2    ($X$)     &37.8  &22.1    &339.6    &15.4  & \\
\cline{1-5}
PP: 2/4    ($X$)     &8.5  &0.2    &352.9    &1.5  & \\
PP: 2/10   ($X$)     &23.0    &2.2    &379.1    &13.2 &  \\
PP: 2/20   ($X$)     &20.0  &10.5    &420.2    &13.2 &  \\ \hline
PP: 20/20  ($\Gamma$)&71.7 &12.2 &206.1 &95.9 \\
PP: 30/20  ($\Gamma$)&41.5 &7.3 &114.5  &111.8 \\
PP: 50/20  ($\Gamma$)&17.9&3.6 &50.4  &126.6 \\
PP: 80/20  ($\Gamma$)&7.7  &1.5 &22.1  &113.6 \\\hline
PP: 3D GaAs &0.0  &{\bf 21.6} &0.0 & {\bf 8.3} \\
$GW$:  3D GaAs &0.0   &{\bf 8.5} &12.6  &{\bf 3.1}    \\
\hline \hline
\end{tabular}
\end{table}

\begin{figure}[!hbp]
\includegraphics[width=0.8\textwidth] {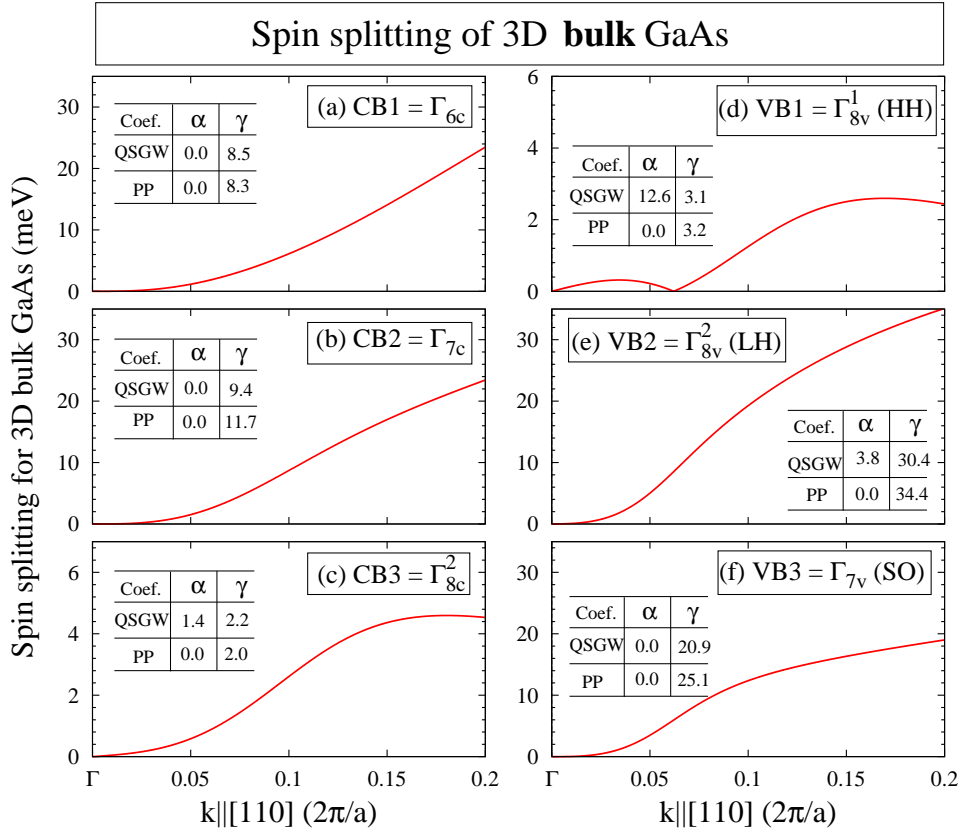}
\caption{\label{fig2} (color online) QS$GW$ predicted SS
of 3D bulk GaAs for the three lowest conduction bands (a) CB1, (b)
CB2, and (c) CB3, and three highest valence bands (d) VB1, (e) VB2,
and (f) VB3. Dresselhaus constants $\alpha_i$ in meV{\AA} and
$\gamma_i$ in eV{\AA}$^3$ predicted by QS$GW$ and screened
pseudopotential fit to QS$GW$, respectively, are given  for each
band in the inset.}
\end{figure}

\begin{figure}[!hbp]
\includegraphics[width=0.7\textwidth] {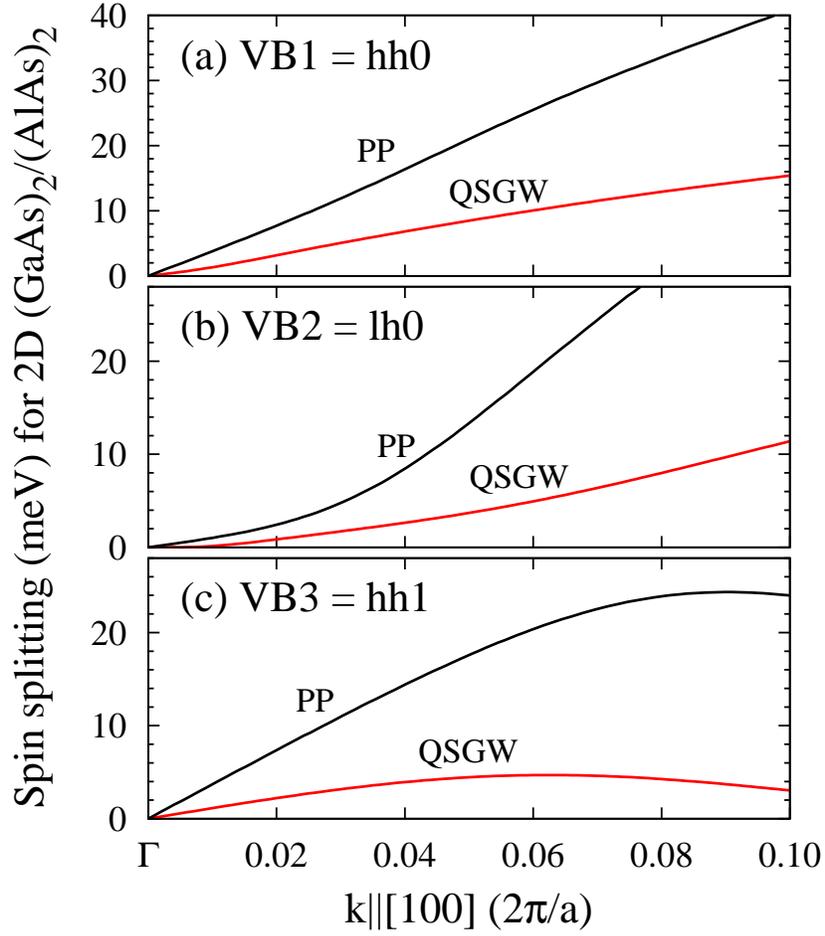}
\caption{\label{fig1} QS$GW$ [Eq. (1)] predicted SS (red
line) of 2D (GaAs)$_2$/(AlAs)$_2$ superlattice for the three highest
valence subbands (a) hh0, (b) lh0, and (c) hh1. The corresponding
SS calculated by screened pseudopotential fit to model
Hamiltonian result \cite{Rashba88} [Eq. (2)] is also given in black
line for each subband.}
\end{figure}

\begin{figure}[!hbp]
\includegraphics[width=0.8\textwidth] {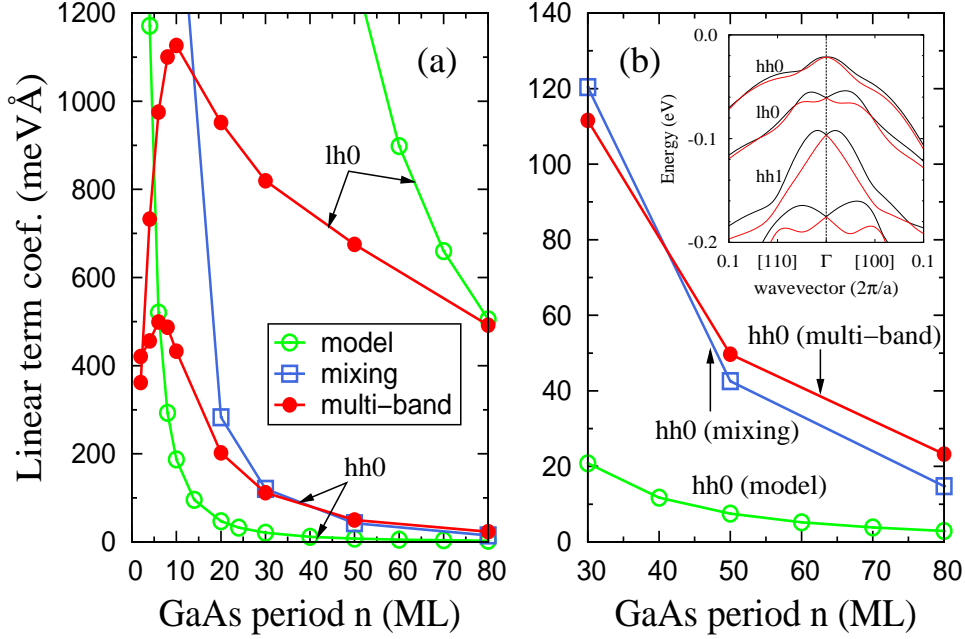}
\caption{\label{fig2} (a) Comparison of linear SS
coefficients $\alpha_{hh0}^{(2D)}$ and $\alpha_{lh0}^{(2D)}$
 calculated from present pseudopotential multi-band approach (same as in Table I; solid
circles), as well as by the Rashba-Sherman model Hamiltonian (open circles), and
mixing approximation [Eq. (4); open squares] for 2D (GaAs)$_n$/(AlAs)$_{20}$ quantum wells. \
(b) Expanded scale for long-period QW's highlighting the comparison of
hh0 subbands for the direct calculation {\it vs} model Hamiltonian and mixing approximation.
Inset shows energy dispersion of valence subbands for 2D
(GaAs)$_n$/(AlAs)$_{20}$ calculated by atomistic pseudopotential approach. The two
spin subbands for each orbit subband are represented by low energy
red line and a high energy black line. }
\end{figure}

\begin{figure}[!hbp]
\includegraphics[width=0.8\textwidth] {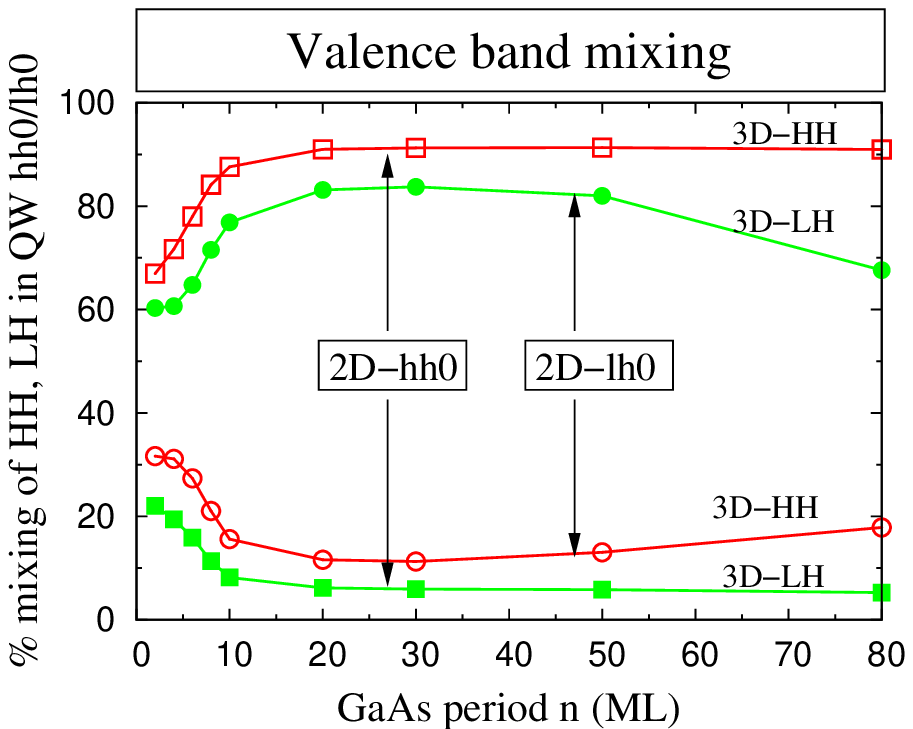}
\caption{\label{fig2} Percent of bulk HH and LH bands mixed into the
2D hh0 and lh0 subbands at zone center calculated by numerical
projection. The results are almost same in a small in-plane $k$
range [e.g., $k<0.01 (2\pi/a)$]. Note that the sum of bulk HH and LH content is less than 100\%. In our atomistic multi-band calculation, the subband ordering from top to bottom is hh0, lh0, hh1, ... for all GaAs period.}
\end{figure}


\begin{references}
\bibitem{Dresselhaus}
G. Dresselhaus, Phys. Rev. {\bf100}, 580 (1955).

\bibitem{Rashba84}
Y.A. Bychkov and E.I. Rashba, J. Phys. C {\bf 17}, 6039 (1984).

\bibitem{Murakami}
S. Murakami, N. Nagaosa, S.C. Zhang, Science {\bf 301}, 1348 (2003).

\bibitem{Brunner_Science}
D. Brunner, {\it et al}, Science {\bf 325}, 70 (2009).

\bibitem{Heiss_PRB}
D. Heiss, S. Schaeck, H. Huebl, M. Bichler, G. Abstreiter, J. J.
Finley, D. V. Bulaev, and Daniel Loss, Phys. Rev. B {\bf76},
241306(R) (2007).

\bibitem{Yogova_PRL}
I. A. Yugova, A. A. Sokolova, D. R. Yakovlev, A. Greilich, D.
Reuter, A. D. Wieck, and M. Bayer, Phys. Rev. Lett. {\bf102}, 167402
(2009)

\bibitem{ZMbook}
{\it Optical Orientation}, edited by B.P. Zakharchenya and F. Meier
(North-Holland, Amsterdam, 1984).

\bibitem{AZ_pss}
A. Zunger, phys. stat. sol. (a) {\bf190}, 467 (2002).

\bibitem{Magri_PRB01}
R. Magri and A. Zunger, Phys. Rev. B {\bf64} 081305(R) (2001).

\bibitem{Cardona_PRB88}
M. Cardona, N.E. Christensen, and G. Fasol, Phys. Rev. B {\bf 38},
1806 (1988).

\bibitem{Ivchenko_PRB}
E. L. Ivchenko, A. Yu. Kaminski, and U. R\"{o}ssler, Phys. Rev. B
{\bf 54}5852 (1996).

\bibitem{Foreman_PRL01}
B.A. Foreman, Phys. Rev. Lett. {\bf86}, 2641 (2001); and references
therein.

\bibitem{Bulaev_PRL}
D. V. Bulaev and D. Loss, Phys. Rev. Lett. {\bf95}, 076805 (2005).

\bibitem{Winkler_book}
R. Winkler, {\it Spin-orbit Coupling Effects in Two-Dimensional
Electron and Hole Systems}, (Springer, 2003).

\bibitem{Rashba88}
E.I. Rashba and E.Ya. Sherman, Phys. Lett. A {\bf129}, 175 (1988).

\bibitem{Bernevig_PRL}
B.A. Bernevig and S.C. Zhang, Phys. Rev. Lett. {\bf 95}, 016801
(2005).

\bibitem{BNL_PRL07}
M.F. Borunda, T. S. Nunner, T. Luck, N. A. Sinitsyn, C. Timm, J.
Wunderlich, T. Jungwirth, A. H. MacDonald, and J. Sinova, Phys. Rev.
Lett. {\bf 99}, 066604 (2007)

\bibitem{SLB09}
D. Stepanenko, M. Lee, G. Burkard, and D. Loss, Phys. Rev. B {\bf
79} 235301 (2009)

\bibitem{OWS08}
O. Olendski, Q. L. Williams, and T. V. Shahbazyan, Phys. Rev. B {\bf
77} 125338 (2008)

\bibitem{SL05} J. Schliemann and D. Loss, Phys. Rev. B {\bf 71} 085308 (2005)

\bibitem{Eppenga_PRB}
R. Eppenga and M.F.H. Schuurmans, Phys. Rev. B {\bf37}, 10923 (R)
(1988).


\bibitem{GW}
M. van Schilfgaarde, T. Kotani, and S. Faleev, Phys. Rev. Lett.
{\bf96}, 226402 (2006).

\bibitem{Chantis_PRL}
A.N. Chantis, M. van Schilfgaarde, and T. Kotani, Phys. Rev. Lett.
{\bf96}, 086405 (2006).

\bibitem{Krich_PRL}
J.J. Krich and B.I. Halperin, Phys. Rev. Lett., {\bf98}, 226802
(2007).

\bibitem{Luo09}
J.W. Luo, G. Bester, and A. Zunger, Phys. Rev. Lett. {\bf102} 056405
(2009).


\bibitem{note1}
The discrepancy of Eq. (4) from the
multi-band calculation for $n< 20$ implies that there are additional
SOI induced interband coupling for small GaAs period $n$, as evidenced by significantly enhanced band mixing (see Fig. 4).











\end{references}
\end{document}